# Two-Dimensional Quantum Dynamics of $O_2$ Dissociative Adsorption on Ag(111)

Yuji Kunisada*, Norihito Sakaguchi

*Center for Advanced Research of Energy and Materials, Faculty of Engineering,*
*Hokkaido University, Sapporo, Hokkaido 060-8628, Japan*
*Email:kunisada@eng.hokudai.ac.jp*

We have investigated the quantum dynamics of $O_2$ dissociative adsorption on a Ag(111) surface. We performed the calculations with a Hamiltonian where the $O_2$ translational motion is perpendicular to the surface and for $O_2$ vibrational energy. We found that dissociative adsorption occurs with an incident translational energy below the expected activation barrier, while the translational-energy dependence for adsorption probabilities is a smooth sigmoid. Thus, there are non-negligible tunneling effects in the dissociative adsorption that are affected by the activation barrier width. Moreover, the incident translational energies at the inflection points of the adsorption probabilities shift lower with increasing in vibrational quantum numbers of the incident $O_2$. Thus, there is significant energy transfer and coupling from vibration to translational motion. The vibrational energy assists the $O_2$ dissociative adsorption *via* a vibrationally assisted sticking effect.

KEYWORDS: density functional theory, oxygen adsorption, dissociative adsorption, potential energy surface, tunneling effect, vibrationally assisted sticking effect

# 1. Introduction

One of the most important goals in material science is a detailed understanding of oxidation. Even though a vast amount of research has been conducted to realize protective oxide layers[1,2], the atomic-scale formation process, which includes the segregation of additional easily oxidizable elements, remains ambiguous. The dissociative adsorption of $O_2$ on solid surfaces is of considerable interest[3,4], because it is a preliminary process in oxidation. Recently, a tunneling effect has been reported for $O_2$ dissociative adsorption on Pt surfaces.[5,6] Tunneling and zero-point energy effects of $O_2$ have also been suggested in other several systems.[7,8] While oxygen atoms are considerably more massive than hydrogen atoms, quantum effects for $O_2$ are not negligible.

The Ag(111) surface is a catalyst for various oxygen-related reactions, such as ethylene epoxidation and partial oxidation of methanol to formaldehyde. The oxidation process and the oxidized phase for Ag(111) surfaces are not fully understood. Therefore, a better understanding of oxidation would lead to a better understanding of the catalytic activity of Ag(111) surfaces. Based on our previous potential energy surface studies,[3,4] we performed quantum dynamics calculations of $O_2$ dissociative adsorption on Ag(111) surfaces by solving the Schrödinger equation for $O_2$. We thus obtain adsorption probabilities as a function of the incident $O_2$ translational energy. Furthermore, we include $O_2$ vibrational energy in the quantum dynamics calculation, and discuss the effect of coupling between translation and vibration.

# 2. Computational Methods

We solved the time-independent Schrödinger equation for $O_2$ moving along the reaction path, using the coupled-channel method.[9-15] The dynamical variables are the $O_2$ center-of-mass (c.m.) distance from the surface $Z$ and the $O_2$ intramolecular bond length $r$. We also neglected surface atom relaxation and the energy dissipation of $O_2$ because they do not affect the direct $O_2$ dissociation process.[16]

To use the coupled channel method, we transform the Hamiltonian in the Cartesian coordinate system into the mass-weighted reaction path coordinate system.[17] The mass-weighted reaction path coordinate $z$ is defined as:

$$z = Z\sqrt{\frac{M}{\mu}}, \tag{1}$$

where $M$ and $\mu$ are the total mass and the reduced mass of $O_2$, respectively. Then, we obtain the Hamiltonian $H$ in the mass-weighted reaction path coordinate system:

$$H = -\frac{\hbar^2}{2M}\frac{\partial^2}{\partial z^2} - \frac{\hbar^2}{2\mu}\frac{\partial^2}{\partial r^2} + V(z,r;X,Y,\theta,\varphi), \tag{2}$$

where $X, Y$ are the positions of the $O_2$ c.m., $\theta, \phi$ are the polar and azimuthal angular orientation of $O_2$ with respect to the surface, and $V(z, r; X, Y, \theta, \varphi)$ is the potential energy of $O_2$ on the Ag(111) surface, respectively. To observe the tunneling effect and coupling between translation and vibration, we focused on translational motion perpendicular to the surface and the vibrational energy. Thus, $X, Y, \theta, \phi$ are parameters. Hereafter, to simplify the expression, we use $V(z, r)$ instead of $V(z, r; X, Y, \theta, \varphi)$. To solve the corresponding Schrödinger equation for dissociative adsorption, it is convenient to make the coordinate transformation $(r, z) \rightarrow (s, v)$, where $(s, v)$ are the reaction path coordinates.[18-20] The position of the $O_2$ c.m. along the reaction path is $s$, and $v$ is a coordinate perpendicular to $s$ at all points along the reaction path. In the coordinate transformation, the Hamiltonian is given by:

$$H = -\frac{\hbar^2}{2\mu}\eta^{-1}\left[\frac{\partial}{\partial s}\eta^{-1}\frac{\partial}{\partial s} + \frac{\partial}{\partial v}\eta\frac{\partial}{\partial v}\right] + V(s,v). \tag{3}$$

The term $\eta$ is the Jacobian for the coordinate transformation, which is given by:

$$\eta = \eta(s,v) = 1 - vC(s), \tag{4}$$

where $C(s)$ is the curvature of the reaction path, which also depends on $X, Y, \theta, \phi$. We used the concept of a local reflection matrix[13] for numerically solving the coupled channel problems. For more details of the calculation we refer to Ref. 13.

We performed coupled-channel calculations by using potential energy surfaces from $O_2$/Ag(111) density functionals.[3,4] Representative adsorption trajectories are shown in Fig. 1, where t, nt, b, f, and h indicate top site, near-top site, bridge site, fcc-hollow site, and hcp-hollow site, respectively. For example, an h–t–f trajectory is where the $O_2$ c.m. is over the top site, and the O–O axis is oriented parallel to the surface along the fcc-hollow and the hcp-hollow sites. The polar orientation of $O_2$ is fixed parallel to the surface. All energies are potential energies with respect to the separate Ag(111) slab and $O_2$ molecule ($E_0 = E_{\text{Ag slab}} + E_{\text{gas-phase oxygen molecule}}$).

## 3. Results and Discussion

In Fig. 2, we plot the functions $V(s,v)$, $C(s)$, and $\hbar\omega(s)$ for $O_2$ dissociative adsorption for the f-nt-f, h-t-f, and b-t-b configurations extracted from previous studies.[3,4] Here, $\hbar\omega(s)$ is the $O_2$ molecular vibration energy that depends on $X, Y, \theta, \phi$. For the f-nt-f, h-t-f, and b-t-b configurations, the respective activation barriers are 1.37, 2.17, 2.18 eV, and the potential energies with metastable dissociative adsorption states are 0.355, 0.552, 1.30 eV, respectively.

The calculated $O_2$ adsorption probabilities for these configurations are shown in Fig. 3. Vibrational states from the ground to the sixth excited state are taken into account. In Fig. 3, we show the results for the ground vibrational state. For all three adsorption configurations, we found that dissociative adsorption occurs with incident translational energy below the activation barrier heights of the corresponding dissociative adsorption. The adsorption probabilities do not reach unity immediately when the incident translational energy exceeds the activation barrier. Thus, the dependence of adsorption probabilities on incident translational energy is a smooth sigmoid. This is due to quantum tunneling. Tunneling effects of $O_2$ and $N_2$ have been reported in $O_2$/Pt[5,6] and $N_2$/Fe systems.[21]

By comparing the results in Fig. 3 for the h-t-f and b-t-b configurations that have very similar activation barrier heights, but different barrier widths (Fig. 2), we found that the gradients of the adsorption probability curve around the inflection point for the h-t-f configuration are smaller than those for the b-t-b configuration. However, the gradients of the adsorption probability curves around the inflection points for the f-nt-f and b-t-b configurations show similar behaviors in Fig. 3, and have almost the same activation barrier widths (Fig. 2) and different activation barrier heights. From these results, we can conclude that the activation barrier width affects the tunneling of $O_2$, while the activation barrier height does not.

In Fig. 4, we show the adsorption probabilities for different incident $O_2$ vibrational states. For the ground vibrational state, the incident translational energies at the inflection points in all three adsorption configurations are slightly lower than the corresponding activation barrier heights. Shifts of the incident translational energies at the inflection points from the activation barrier heights are 83.9, 84.8, and 83.9 meV for the f-nt-f, h-t-f, and b-t-b configurations, respectively. We define $\hbar\Delta\omega$ by:

$$\hbar\Delta\omega = \hbar\omega_{O_2} - \hbar\omega_{2O/Ag(111)} = \hbar\omega(-\infty) - \hbar\omega(+\infty). \tag{5}$$

Where $\hbar\omega_{O_2}$ is 192 meV, and $\hbar\omega_{2O/Ag(111)}$ is 24.2, 22.4, and 24.2 meV, for the f-nt-f, h-t-f, and b-t-b configurations, respectively. These shifts correspond to $\hbar\Delta\omega/2$, which is the energy difference between the incident and final $O_2$ vibrational ground states. With increasing incident vibrational quantum number, the incident translational energies at the inflection points for all three adsorption configurations shift to lower values. These shifts in inflection point are equal to integral multiples of $\hbar\Delta\omega$, and suggest that vibrational energy transfer to translational motion occurs *via* vibrational-translational coupling. Therefore, the $O_2$ vibrational energy can assist dissociative adsorption because of the vibrationally assisted sticking (VAS) effect.[22-24] For a significant VAS effect, a finite curvature $C(s)$ of the reaction path is required below the activation barrier[20,25], *i.e.,* a "*late-barrier*" reaction. In Fig. 2, we found that the peaks of the curvature $C(s)$ occur below the activation barrier in all three configurations. Thus, an effective VAS effect occurs in $O_2$ dissociative adsorption on Ag(111).

We also considered the effect of $O_2$ gas temperature on adsorption probabilities. The dependence, along with incident translational energies, can be calculated with the Boltzmann distribution function:

$$P(E_t, T) = \frac{\sum_{v=0}^{6} \sum_{J=1,\text{odd}}^{79} (2J+1) S_v(E_t) \exp\left(-\frac{E_t + E_v + E_J}{k_B T}\right)}{\sum_{v=0}^{6} \sum_{J=1,\text{odd}}^{79} (2J+1) \exp\left(-\frac{E_t + E_v + E_J}{k_B T}\right)}, \tag{6}$$

$$E_v = \hbar\omega_{O_2}\left(v + \frac{1}{2}\right), \tag{7}$$

$$E_J = BJ(J+1), \tag{8}$$

where T, $E_t$, $E_v$, $E_J$, and $S_v(E_t)$ are the incident $O_2$ gas temperature, translational energy, vibrational energy for quantum number $v$, rotational energy for quantum number $J$, and the adsorption probabilities for the corresponding vibrational states, respectively. *B* is the $O_2$ rotational constant (0.18 meV). To understand the effects of vibrational motion on adsorption probabilities, we assumed that the $O_2$ rotational energy affects only the Boltzmann distribution. We only considered adsorption probabilities for the f-nt-f configuration, and ground- to fortieth-excited rotational states. We note that $^{16}O_2$ can have only odd azimuthal rotational quantum numbers because of the symmetric exchange of nuclei. The coefficient $2J+1$ is because of the rotational degeneracy. In Fig. 5, we show the adsorption probabilities for different incident $O_2$ translational energies and gas temperatures. For the same incident $O_2$ translational energy, we found that the adsorption probabilities increase with the gas temperature. Similar trends have been reported experimentally for the $D_2$/Cu(111) system.[24] In the $D_2$/Cu(111) system, the increase of adsorption probabilities with the gas temperature rise has been observed in the low incident $D_2$ translational energy region. The origin of these trends can be comprehended as follows. With increasing the gas temperature, the molecular vibrational states are thermally excited. Thus, even in the incident translational energy region less than the activation barriers, these molecules can overcome the corresponding activation barriers with the assistance from the excited molecular vibration. From this point, this experimental result is a direct observation of the VAS effect.

4. Conclusion

We have investigated the quantum dynamics of $O_2$ dissociative adsorption on an Ag(111) surface, by solving the Schrödinger equation for $O_2$ moving along the reaction path. The Hamiltonian was formulated for $O_2$ translational motion perpendicular to the surface and for $O_2$ vibration. The calculated adsorption probabilities were for the f-nt-f, h-t-f, and b-t-b configurations. We found that dissociative adsorption occurs with an incident translational energy below the activation barrier heights, and that the dependence of adsorption probabilities on incident translational energy is a smooth sigmoid. Thus, we concluded that there exists a non-negligible tunneling effect. Moreover, we revealed that the widths of the activation barriers affect the $O_2$ tunneling, while the activation barrier heights do not. We also found that the incident translational energies at the inflection points decrease with increasing vibrational quantum number. This is due to coupling and transfer of vibrational energy to translational motion. We also revealed an increase in adsorption probabilities with $O_2$ gas temperature. Thus, we concluded that vibrational energy can assist the dissociative adsorption *via* the VAS effect. We can understand the oxidation of Ag(111) surfaces in greater detail by considering even more degrees-of-freedom in the quantum dynamics calculations. Previous studies have reported that rotational motion can also affect adsorption probabilities.[26,27] An ongoing study involves molecular rotation and Ag(111) surface vibrations. These calculations can give us more insight into ways to control the catalytic activities of the O/Ag system.

**Acknowledgments**

Calculations were performed with the computer facilities of the Information Initiative Center (Hokkaido University) and the ISSP Super Computer Center (The University of Tokyo). The authors are thankful to H. Kasai of Osaka University for helpful discussions

**Fig. 1** Reaction pathways. t, nt, b, f, h are top, near-top, bridge, fcc-hollow, hcp-hollow sites, respectively. Black and white circles indicate positions of O and Ag atoms, respectively, and the parallelogram is the fcc (2x2) supercell of the Ag(111) surface.

**Fig. 2** Potential energy, curvature, and $O_2$ vibrational energy along the reaction path in (a) f-nt-f, (b) h-t-f, and (c) b-t-b configurations.

**Fig. 3** Adsorption probabilities with the initial $O_2$ vibrational state in its ground state in f-nt-f, h-t-f, and b-t-b configurations.

**Fig. 4** Adsorption probabilities with the initial $O_2$ vibrational state in its ground to third excited states in (a) f-nt-f, (b) h-t-f, and (c) b-t-b configurations. Here, *n* refers to the vibrational quantum numbers.

**Fig. 5** $O_2$ gas temperature dependence for adsorption probabilities in f-nt-f configurations.

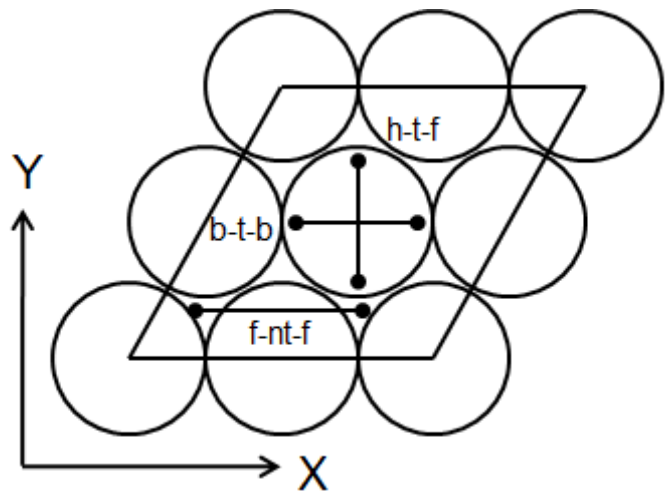

Fig. 1

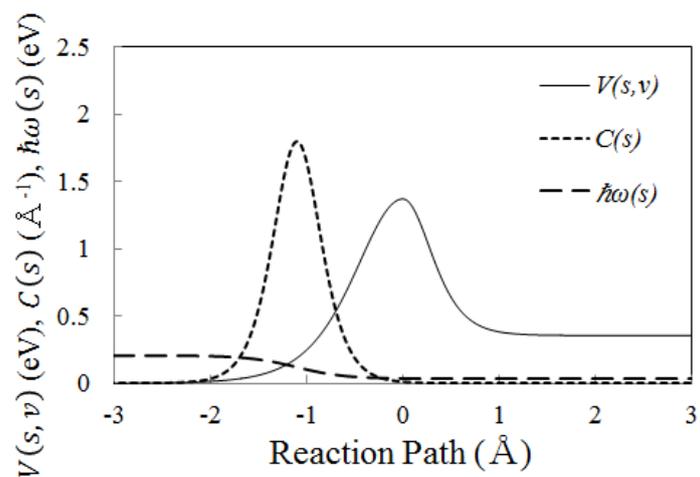

(a)

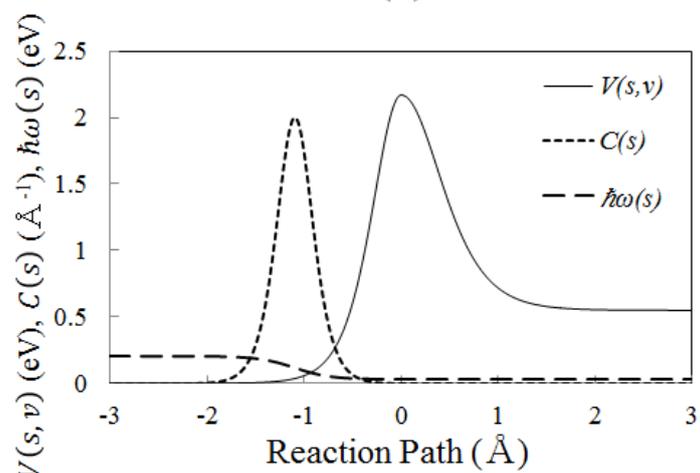

(b)

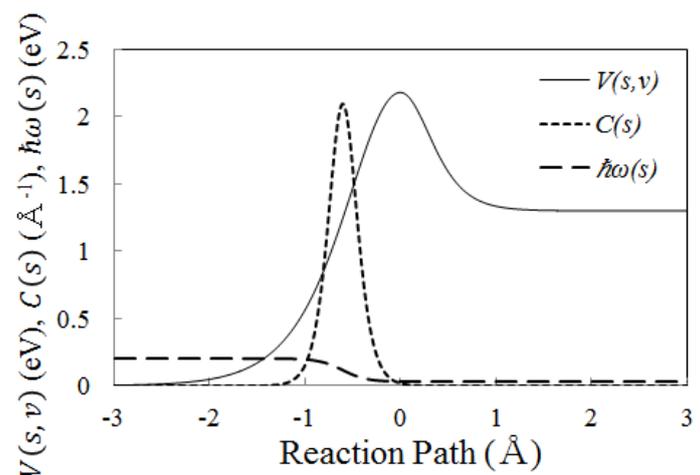

(c)

Fig. 2

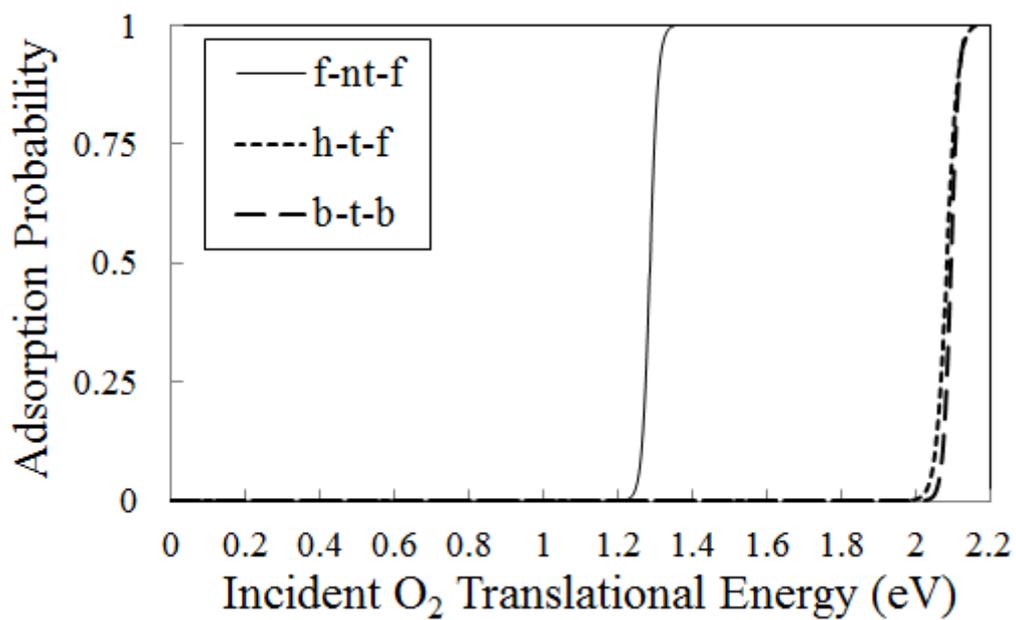

Fig. 3

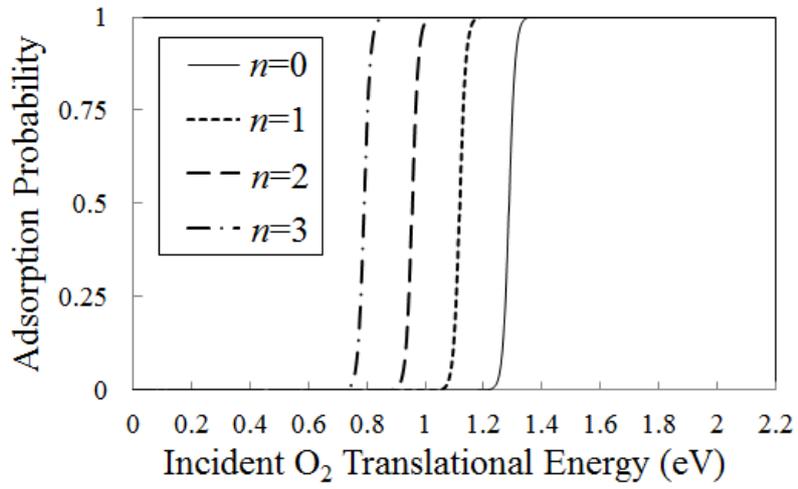

(a)

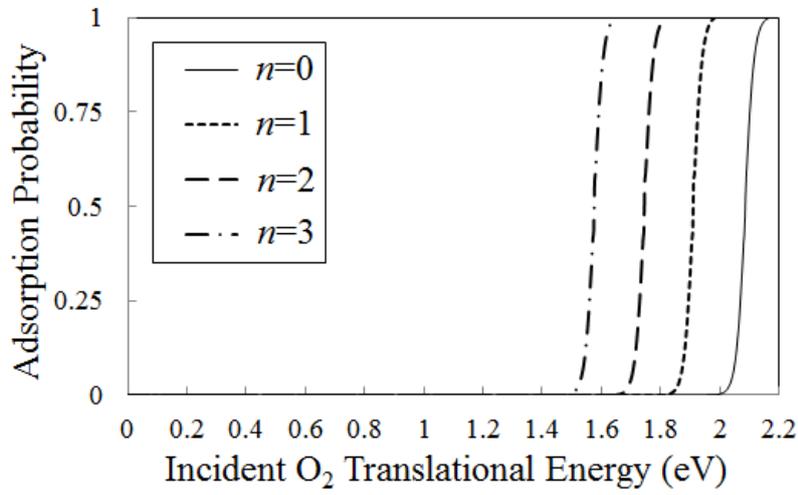

(b)

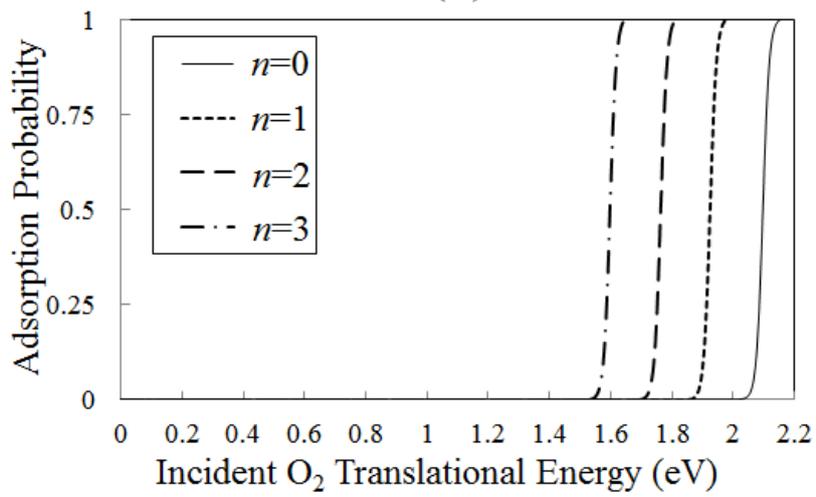

(c)

Fig. 4

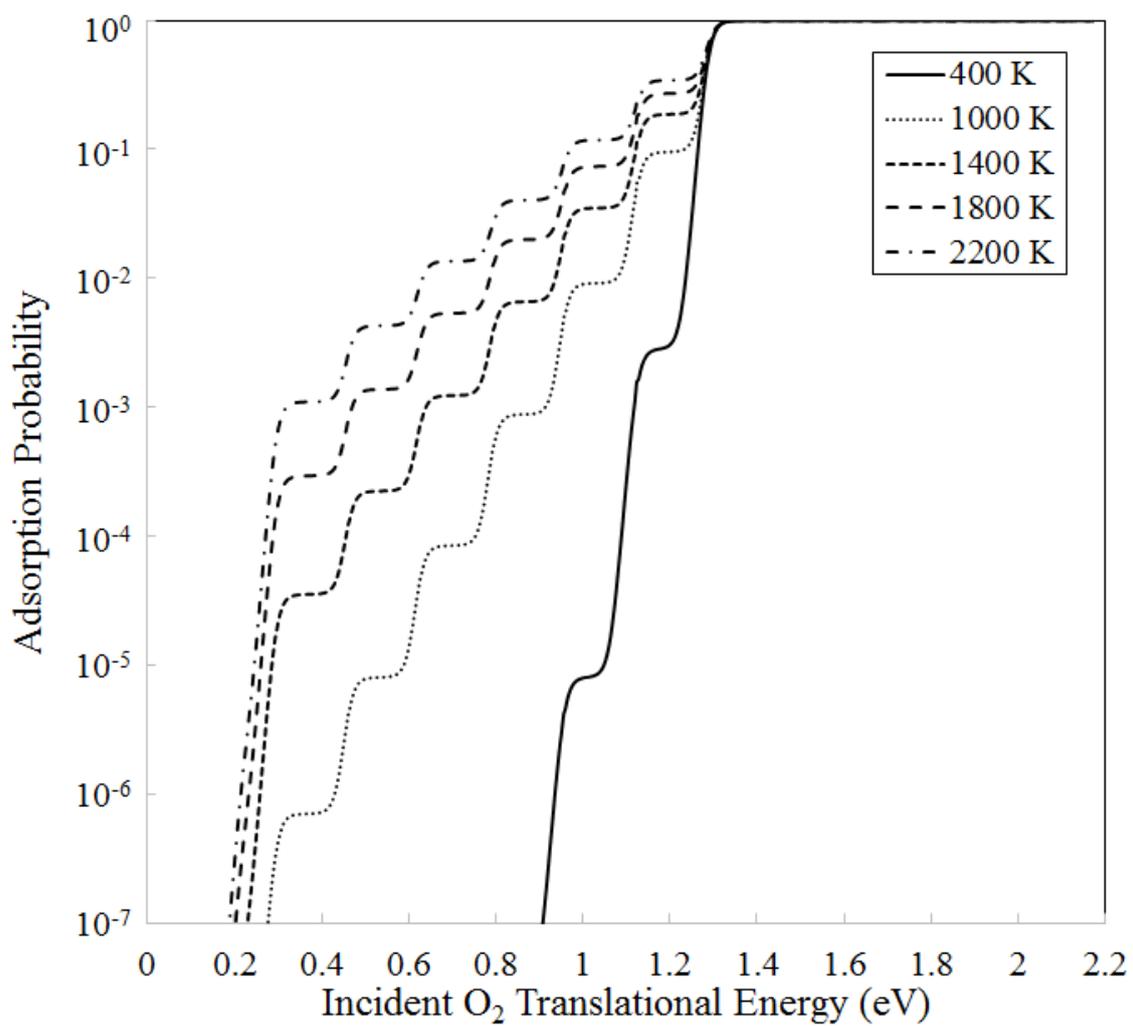

Fig. 5